Metalorganic chemical-vapor deposition of high-reflectance III-nitride distributed Bragg reflectors on Si substrates

M.A. Mastro, R.T. Holm, N.D. Bassim, D.K. Gaskill, J.C. Culbertson, M. Fatemi, C.R. Eddy Jr., R.L. Henry, M.E. Twigg
U.S. Naval Research Laboratory, 4555 Overlook Ave., SW, Washington, D.C. 20375

**Abstract**
High-reflectance group III-nitride distributed Bragg reflectors (DBRs) were deposited by MOCVD on Si (111) substrates. A reflectance greater than 96% was demonstrated for the first time for an AlN/GaN DBR with a stop-band centered in the blue-green range of the visible spectrum. Crack-free GaN cap layers were grown on the DBR structures to demonstrate the opportunity to build III-nitride optoelectronic devices in this material. The DBR structure was under significant strain due to growth on a mismatched substrate although the GaN cap layer was shown to be strain free.

**Introduction**

The escalating III-nitride based LED market is projected to exceed four billion dollars by 2008 [1]. This prediction is based on the assumption of continuing decrease in per die LED cost with a simultaneous increase in LED external efficiency. In fact, penetration into the general white illumination market is based on prediction of a 10x increase in lumen/$ over next twenty years [2]. The cost of standard brightness LEDs will be partially addressed by a shift to low-cost manufacturing centers while the gain in external efficiency for high-brightness LEDs will come from complicated techniques such as flip-chip die shaping for improved light extraction [3].

Presently, the standard and high brightness blue, green and white LED markets are almost entirely supplied by III-nitride optoelectronic devices grown by MOCVD on sapphire and SiC substrates [4,5]. These substrates represent a significant fraction of the overall cost to fabricate a LED die. A promising alternative is to substitute Si as the substrate, which, on a per square inch basis, has at most one-tenth the cost of sapphire and one-hundredth the cost of SiC. Additionally, large size (2 to 12 inch diameter), high quality Si substrates are readily available due to the maturity of the silicon industry.

Epitaxy of high-quality group III-nitride films on Si is difficult due to the large lattice and thermal expansion coefficient (TEC) mismatch between the film and the substrate. A number of groups have implemented a thin layer of AlN to compensate for the lattice parameter and TEC mismatch between the silicon substrate and active gallium nitride (GaN) device layer [6]. Still, the lattice mismatch between AlN and Si (111) is 23.5% due to the lattice spacing of 0.3112 nm and 0.384 nm, respectively. Recently, an examination by Bourret et al. found that 4:5 AlN:Si coincident site lattices can form at the interface with the (111) plane, which would alter the mismatch to -1.2% (compressive) [7]. In practice, the initiation of AlN eptiaxy on Si entails nucleation and coalescence of 3-D islands emblematic of a Volmer-Weber (VW) type growth [8,9]. The lattice mismatch is usually absorbed within the first few monolayers by misfit dislocations. The tensile stress generated at growth temperature is characteristic of adjacent islands zippering together to reduce elastic strain [10]. Furthermore, the intrinsic tensile stress generated during growth is further enhanced during cool down due to the large TEC mismatch with macro-crack formation customary for GaN films thicker than 1 μm [11].

A number of approaches have been developed to introduce a compressive stress into the structure during growth to offset the large TEC mismatch tensile stress, including superlattices (SLs) of GaN and AlGaN [12,13]. The negative lattice mismatch for GaN grown on AlGaN (or AlN) compresses the GaN film during 2-D epitaxy. Similarly, Nitronex employs an AlN (~0.3 μm)/graded-AlGaN (~0.5 μm) structure [14,15] to subsequently grow crack-free GaN/AlGaN for the commercial production of microwave transistors. Although several groups have demonstrated GaN-based LEDs on Si substrates [16-18], there are no reports of volume production of GaN LEDs deposited on Si substrates.

The principal requirement to improve the efficiency of GaN-based optoelectronic devices on Si is to reduce the high defect and crack densities in the III-nitride film despite the large lattice and thermal expansion mismatch with the substrate. The second limitation to the efficiency of III-nitride optoelectronic devices is optical absorption in the UV and visible spectrum by the opaque Si substrate. Inserting a high-reflectance distributed Bragg reflector (DBR) between the active portion of the device and the substrate would reflect the emitted light away from the opaque Si, thus approximately doubling the efficiency of the device. Additionally, the DBR structure described in this document counters the TEC mismatch tensile strain, in addition to filtering the dislocations that propagate from the film/substrate interface that can be deleterious to device performance.

**Experimental**

This letter reports on metalorganic chemical vapor deposition (MOCVD) of an GaN nano-rods on ErCl3 seeded a-plane sapphire substrates. Erbium chloride (s) was dissolved in deionized water then sprayed or spun onto the two-inch sapphire substrate. Proper application of the liquid in limited volume prevented accumulation of excess seed on the substrate. The substrate was immediately heated to approximately 60 °C to drive off any excess water. Growth was carried out in a modified vertical impinging flow chemical vapor deposition reactor. After loading, the wafers were immediately ramped to the growth temperature. A $H_2$ carrier gas in a 50 torr environment is used during the temperature ramp to drive off any residual water on the substrate surface. An Ga seed layer was deposited for 2 sec prior to the onset of $NH_3$ flow to protect the Er seed from nitridation. The GaN structures were deposited at 700 to 950 °C at 50 Torr for 5 to 30 min.

Structural characterization was performed with a Hitachi H-9000 top-entry transmission electron microscope operated at 300 kV and a Panalytical X'pert x-ray diffraction (XRD) system.

**Results**

In a previous article, the authors reported on a high-reflectance III-nitride DBR structure with the stop-band centered in the blue portion of the visible spectrum [20]. The stop-band for the DBR structure can be tuned to any point in the UV or visible by adjusted the thickness of the AlN and $Al_xGa_{1-x}N$ layers in the SL. The reflectance displayed in Fig. 1 is for a DBR with a stopband centered in the blue-green at 2.5 eV (495 nm) with 60 nm AlN and 51 nm GaN layers. The measured reflectance of 96.3% slightly exceeded the calculated reflectance of 95.7% which may indicate a slight error in the refractive index values [21] used in the transfer Matrix calculation [22]. This is the highest reflectance reported for a III-nitride DBR on Si with a stop-band centered in the blue-green region of the visible spectrum.

Raman characterization was used to compare the strain state of a 5x AlN/GaN DBR versus a 7x AlN/GaN DBR structure with a 500 nm GaN cap layer (Fig. 2). A shift to shorter wavenumbers indicates a stretching of the III-nitride bonds due to a tensile stress in the layer. The wavenumber shift to stress relation was reported to be approximately 5 $cm^{-1}$/GPa for AlN [23] and 2.9 $cm^{-1}$/GPa for GaN [24]. The 5x DBR displayed an AlN peak at 639.5 $cm^{-1}$, which corresponds to a tensile stress of 3.28 GPa and a GaN peak at 570 $cm^{-1}$, which corresponds to a compressive stress of 690 MPa. Comparatively, the 7x DBR with a 500 nm cap displayed an AlN peak at 635.7 cm$^{-1}$, which corresponds to a tensile stress of 4.04 GPa while the GaN layers in the DBR were obscured by the thick GaN cap layer, which was effectively unstrained.

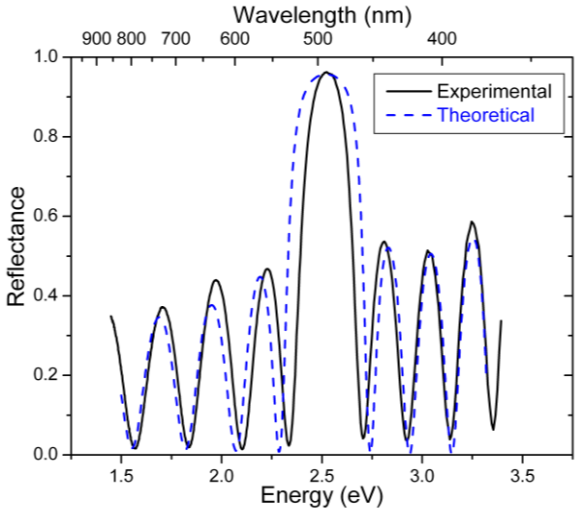

Fig. 1 Optical reflectance of a 9x AlN/GaN DBR structure on a Si (111) substrate displaying greater than 96% reflectivity with the stop-band centered in the blue-green region of the visible spectrum.

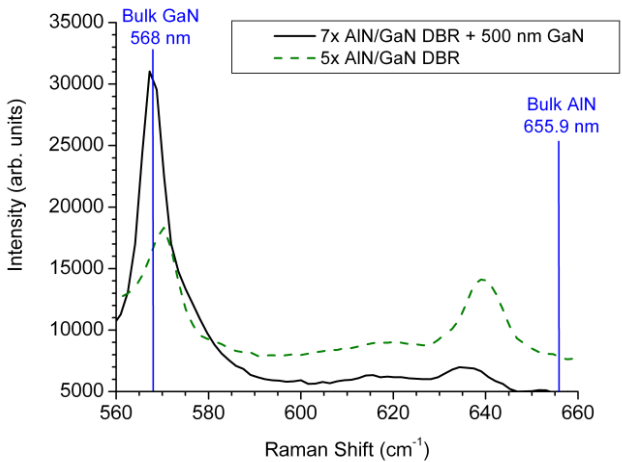

Fig. 2. Raman shift of the E2 phonon for nominally uncracked 5x DBR and 500 nm GaN / 7x DBR structures deposited on Si (111) substrates by MOCVD.

A reciprocal space map near the [1 0 -1 5] asymmetric reflection provides insight into the strain state of III-nitride thin films. In Fig. 3, the signal intensity is plotted against the deviation in reciprocal lattice space relative (approximately) to the AlN [1 0 -1 5] reflection. The AlN and GaN reflections for the 5x DBR along a constant $Q_{1010}$ line indicates the AlN and GaN layers were grown pseudomorphically. The 7x DBR with a GaN cap layer displays a strong [1 0 -1 5] GaN reflection due to the thickness (i.e., volume) of the material and the improvement in material quality with thickness. The reflection intensity from the GaN cap layer is approximately an order of magnitude larger than the GaN reflection from the 5x DBR. Thus, the intensity of GaN cap reflection obscures the reflection signal of the GaN layers in the DBR. The GaN reflection of the cap layer is shifted in $Q_{1010}$ relative to the AlN reflection, indicating that the GaN cap is in a relaxed state similar to what was observed in the Raman characterization described above.

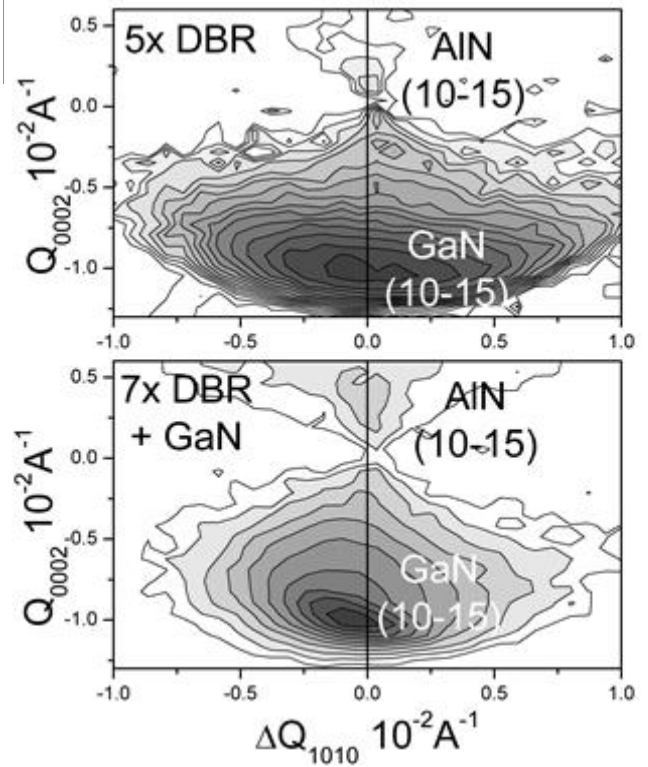

Fig. 3. Reciprocal space map of the asymmetric reflection [1 0 -1 5] for a 5x AlN/GaN DBR (top) and a 7x AlN/GaN DBR with a 500 nm GaN cap layer (bottom).

To integrate a III-nitride optoelectronic device into this GaN/DBR/Si structure, it is first necessary to demonstrate the crack-free and essentially relaxed state of the GaN cap. The subsequent hurdle is reducing the dislocation density to a sufficient level to ensure adequate internal efficiency for a device fabricated in this material. Standard brightness III-nitride LEDs, which employ InGaN can efficiently emit blue (or green) light despite dislocation densities on the order of $10^9$ cm$^{-2}$ [4]. Spontaneous compositional and spatial fluctuations in the InGaN active region creates quantum dot like states that isolate the radiative emission centers from the dislocation non-radiative recombination centers [5].

The lattice mismatch between AlN and Si generates, during the initial stages of AlN island growth, a defect density greater than $10^{12}$ cm$^{-2}$ at the interface. Dadgar and coworkers grew a thick GaN buffer layer on this thin AlN layer to reduce the dislocation density to less than $10^9$ cm$^{-2}$ [17,18]. Similarly, an optimized growth on sapphire or SiC with a thick GaN buffer layer can reduce the dislocation density to the same level [19]. The alternative presented in this paper is to use the SL structure to improve the structural quality of the film in addition to its optical function as a DBR. Fig. 4 displays a cross-sectional TEM micrograph of a nominally crack-free 7x AlN/Al$_{0.05}$Ga$_{0.95}$N DBR with a 500 nm GaN cap on a Si (111) substrate. The dislocation density drops by approximately two to three orders of magnitude through the GaN/SL structure.

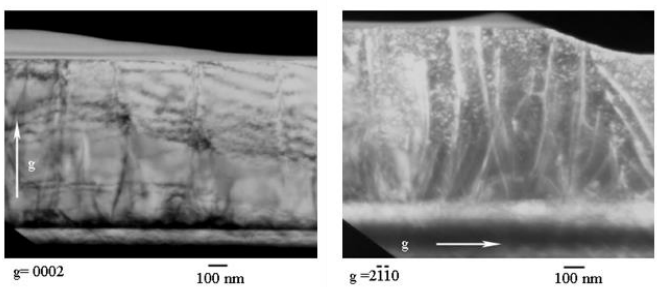

Fig. 4. Electron micrographs of the 500 nm GaN cap layer grown on a nominally crack-free 7x AlN/Al$_{0.05}$Ga$_{0.95}$N DBR on a Si (111) substrate. On the (left), the dark-field technique found a decrease in screw/mixed type dislocations from greater than $10^{12}$ cm$^{-2}$ at SL/substrate interface to approximately $6 \times 10^9$ cm$^{-2}$ at surface. On the (right), the weak beam technique found a decrease in edge type dislocations from greater than $10^{12}$ cm$^{-2}$ at SL/substrate interface to approximately $1.6 \times 10^{10}$ cm$^{-2}$ at surface.

An AFM scan is displayed in Fig. 5 of the surface of a 500 nm GaN cap layer on a 7x DBR. The surface is smooth with a RMS roughness of 0.697 nm. Despite the overall smoothness of the surface, a moderate level of screw type dislocations are observed in the AFM scan of the GaN surface as was also evident in the electron micrograph in Fig. 4.

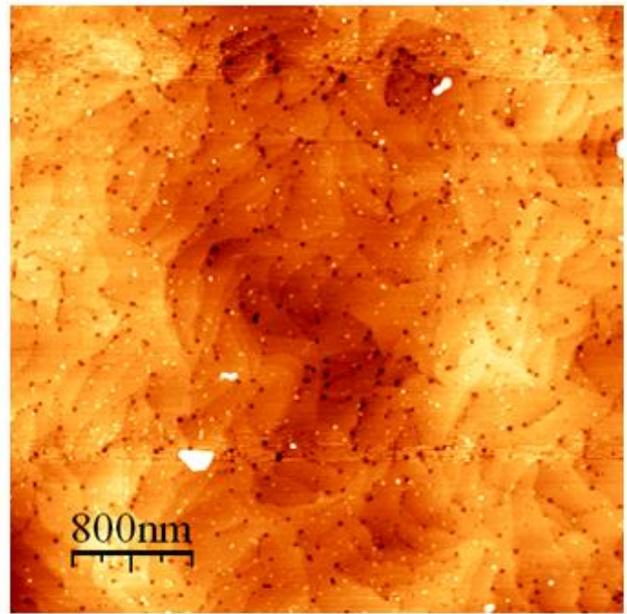

Fig. 5. Tapping mode AFM scan of the surface of a 500 nm GaN / 7x AlN/Al$_{0.05}$Ga$_{0.95}$N DBR / Si (111) structure. The full range of the vertical height in this 4 μm x 4 μm scan is approximately 3.5 nm.

## Conclusion

The ability to place the III-nitride DBR directly on Si allows the design of III-nitride optoelectronic devices without light absorption in the opaque substrate and enables potentially a 10x cost reduction. In this paper, an AlN/GaN DBR displayed greater than 96% reflectance in the blue-green spectrum. The high index contrast at the AlN/Si interface enhanced the reflectance of the AlN/GaN DBR. This SL served the added purpose of counteracting the large TEC mismatch tensile stress. Additionally, the dislocation density decreased by three orders of magnitude through the crack-free DBR / 500 nm GaN structure.

**Acknowledgements**

Research at NRL is supported by ONR; support for Mastro and Bassim was partially provided by the ASEE.